\documentclass[preprint,12pt]{elsarticle}

\usepackage{graphicx}
\usepackage{epsfig}

\newcommand{\be}{\begin{eqnarray}}
\newcommand{\ee}{\end{eqnarray}}

\usepackage{amssymb}

\begin{document}

\begin{frontmatter}


\title{Forward production of beauty baryons in $pp$ collisions at LHC}


\author{ V.A.Bednyakov, G.I.Lykasov, V.V.Lyubushkin}

\address{JINR, Dubna, 141980, Moscow region, Russia}
\begin{abstract}
The production of charmed and beauty baryons
in proton-proton collisions at high
energies is analyzed within the modified quark-gluon string model.
 We present some predictions for the experiments on the forward
beauty baryon production in $pp$ collisions at LHC energies. This analysis
allows us to find useful information on the Regge trajectories of 
the heavy ($b{\bar b}$) mesons and the sea beauty quark distributions in 
the proton.
\end{abstract}

\end{frontmatter}


\section{Introduction}
\label{1}

Various approaches of perturbative QCD including the
 next-to-leading order calculations (NLO QCD) have been applied 
 to analyze the production of heavy flavour mesons and baryons 
in hard $pp$ collisions, see for example \cite{Nasson,Kniehl3} 
and references therein. However, it is impossible to use these approaches
to study the hadron production at small values of the transfer because 
the running QCD constant $\alpha_s$ can be large. A rather successful
description of various characteristics of heavy-quark hadroproduction
processes can be obtained by using the approach for describing the soft 
hadron-nucleon, hadron-nucleus and nucleus-nucleus interactions
at high energies  based on the topological 
$1/N$ expansion in QCD \cite{tHooft:1974,Veneziano:1975}, 
where $N$ is the number of flavours or colours, and the closely related string and 
colored-tube models \cite{Casher1,Gurvich}.
There are successful phenomenological approaches for describing the soft hadron-nucleon,
hadron-nucleus and nucleus-nucleus interactions at high energies 
based on the Regge theory and the $1/N$ 
expansion in QCD, for example the quark-gluon string model (QGSM) \cite{kaid1},
the VENUS model \cite{Werner:1993}, the dual parton model (DPM) \cite{Capella:1994}.

The main components of the QGSM and the DPM are the quark distributions in a hadron
and the the fragmentation functions (FF) describing quark fragmentation into hadrons. 
These are expressed in terms of intercepts of the Regge trajectories $\alpha_R(0)$. The 
largest uncertainty in the calculations of the cross sections for the yields of heavy flavours in 
these models is mainly due to the absence of any reliable information on the transfer $t$ dependence
of the Regge trajectories of heavy quarkonia ($Q{\bar Q}$). Assuming linearity of the   
($Q{\bar Q}$) trajectories, the intercepts turn out to be low, $\alpha_\psi(0)=-2.2$,
$\alpha_\Upsilon(0)=-8, -16$, and so the contribution of the peripheral mechanism decreases very 
rapidly with increasing quark mass, Accordingly, the finding of the $t$-dependence for
$\alpha_{(Q{\bar Q})}(t)$ in the region $0\leq t\leq M^2_{(Q{\bar Q})}$ and estimations of their 
intercepts become especially important.     

In this paper we investigate the open charm and beauty baryon production
in $pp$ collisions at LHC energies and small $p_t$ within the QGSM to find
the information on the Regge trajectories of the heavy ($c{\bar c}$) and ($b{\bar b}$) mesons,
the FF of all the quarks and diquarks to heavy flavour hadrons  
and the sea beauty quark distribution in the proton. 
 We also present some predictions for the future LHC experiments on the forward
$\Lambda_b$ production in $pp$ collisions.

\section{General formalism for hadron production in $pp$ collisions within QGSM}
Let us briefly present the scheme of the analysis of the hadron production in the $pp$ 
collisions within the QGSM including the transverse motion of quarks and diquarks in
colliding protons \cite{kaid2}.
As is known, the cylinder type
graphs for the $pp$ collision presented in Fig.1 make
the main contribution to this process \cite{kaid1}. 
The left diagram of Fig.1, the so-called
one-cylinder graph, corresponds to the case where two colourless
strings are formed between the quark/diquark ($q/qq$) and the
diquark/quark ($qq/q$) in colliding protons; then, after their
breakup, $q{\bar q}$ pairs are created and fragmentated to a hadron,
for example, $D$ meson. The right diagram of Fig.1, the
so-called multicylinder graph, corresponds to creation of the same
two colourless strings and many strings between sea
quarks/antiquarks $q/{\bar q}$ and sea antiquarks/quarks ${\bar
q}/q$ in the colliding protons.
\begin{figure}[htb]
\vspace{9pt}
\includegraphics[scale=1.0]{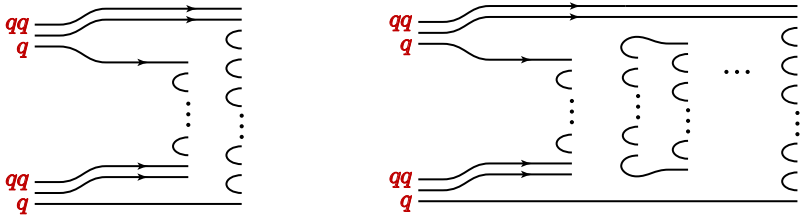}
\caption{The one-cylinder graph (left) and the multicylinder 
graph (right) for the inclusive $p p\rightarrow h X$ process.}
\label{Fig.1}
\end{figure}
The general form for the invariant inclusive hadron spectrum
within the QGSM is \cite{Capella:1994,kaid2}
\begin{eqnarray}
E\frac{d\sigma}{d^3{\bf p}}\equiv
\frac{2E^*}{\pi\sqrt{s}}\frac{d\sigma}{d x d p_t^2}=
\sum_{n=1}^\infty \sigma_n(s)\phi_n(x,p_t)~, 
\label{def:invsp}
\end{eqnarray}
where $E,{\bf p}$ are the energy and the three-momentum of the
produced hadron $h$ in the laboratory system (l.s.); 
$E^*,s$ are the energy of $h$ and the square of the initial energy in the
c.m.s of $pp$; $x,p_t$ are the Feynman variable and the transverse
momentum of $h$; $\sigma_n$ is the cross section for production of
the $n$-Pomeron chain (or $2n$ quark-antiquark strings) decaying
into hadrons, calculated within the quasi-``eikonal approximation''
\cite{Ter-Mart}. Actually, the function $\phi_n(x,p_t)$ is the convolution
of the quark (diquark) distributions and the FF, see the details 
in \cite{kaid1} and \cite{Capella:1994,kaid2}.
\be
\phi^{p p}_n(x)=F_{qq}^{(n)}(x_+)F_{q_v}^{(n)}(x_{-})+
F_{q_v}^{(n)}(x_+)F_{qq}^{(n)}(x_-)+
2(n-1)F_{q_s}^{(n)}(x_+)F_{{\bar q}_s}^{(n)}(x_-)~,
\ee
where
$x_{\pm}=\frac{1}{2}(\sqrt{x^2+x_t^2}\pm x)$~, 
and
\be
F_\tau^{(n)}(x_\pm)=\int_{x_\pm}^1 dx_1 f_\tau^{(n)}(x_1)G_{\tau\rightarrow h}
\left(\frac{x_\pm}{x_1}\right)~,
\label{def:Ftaux}
\ee
 Here $\tau$ means the flavour of the valence (or sea) quark or diquark, $f_\tau^{(n)}(x_1)$
is the quark distribution function depending on the longitudinal momentum fraction $x_1$  
 in the $n$-Pomeron chain; $G_{\tau\rightarrow h}(z)=
z D_{\tau\rightarrow h}(z)$, $ D_{\tau\rightarrow h}(z)$ is the FF of a quark (antiquark) or 
diquark of flavour $\tau$ into a hadron $h$ (charmed hadron in our case);
$\sigma_n$ is the cross section for production of the $n$-Pomeron chain (or $2n$ quark-antiquark strings)
 decaying
into hadrons, calculated within the ``eikonal approximation'' \cite{kaid1}.

To calculate the interaction function $\phi_n(x,p_t)$ we have to know all the quark (diquak)
distribution functions in the $n$th Pomeron chain and the FF. 
They are constructed within the QGSM using the knowledge of the secondary Regge trajectories,
see details in \cite{kaid1,kaid2}. 
Note that the modified QGSM including the internal longitudinal and transverse motion of quarks 
(diquarks) in the initial nucleon has been successfully applied \cite{LAS}-\cite{LLB:09} 
to describe the production of heavy flavour hadrons produced in $p-p$ and $p-{\bar p}$ collisions
at not large $p_t$. 

\section{Heavy baryon production within QGSM}
\subsection{Sea charm and beauty quark distribution in proton}
According to the QGSM, see for example \cite{LAS} and references therein, the distribution 
of $c({\bar c})$ quarks relevant to the $n$th Pomeron 
chain (Fig.1(right)) is, see for example \cite{LAS} and references therein,
\begin{eqnarray}
f_{c({\bar c})}^{(n)}(x) = C_{c({\bar c})}^{(n)}\delta_{c({\bar c})}
x^{a_{cn}} 
(1-x)^{g_{cn}}
\quad
\label{def:fc}
\end{eqnarray}
where $a_{cn}=-\alpha_\psi(0)$, 
$g_{cn}=2(\alpha_\rho(0)-\alpha_B(0))-\alpha_\psi(0))+n-1$;
$\delta_{c({\bar c})}$ is the weight of charmed pairs in the quark sea, 
$C_{c({\bar c})}^{(n)}$
is the normalization coefficient \cite{kaid2},
 $\alpha_\psi(0)$ is the intercept of the $\psi$- Regge trajectory.
Its value can be $-2.18$ assuming that this trajectory $\alpha_\psi(t)$ 
is linear and the intercept and the slope $\alpha_\psi^\prime(0)$ can be
determined by drawing the trajectory through the $J/\Psi$-meson mass 
$m_{J/\Psi}\simeq 3.1$ GeV and the $\chi$-meson mass $m_\chi=3.554$ GeV 
\cite{Boresk-Kaid:1983}. Assuming that the $\psi$- Regge trajectory is 
nonlinear, one can get  $\alpha_\psi(0)\simeq 0$, which follows from the
QCD-based model analysis \cite{Sergeenko:94,Likhoded:06}. 
The  distribution of $b({\bar b})$ quarks in the $n$th Pomeron 
chain (Fig.1(right)) has the similar form  
\begin{eqnarray}
f_{b({\bar b})}^{(n)}(x) = C_{b({\bar b})}^{(n)}\delta_{b({\bar b})}
x^{a_{bn}}
(1-x)^{g_{bn}}
\quad
\label{def:fb}
\end{eqnarray}
where $a_{bn}=-\alpha_\Upsilon(0)$, $g_{bn}=2(\alpha_\rho(0)-\alpha_B(0))-\alpha_\Upsilon(0))+n-1$; 
$\alpha_\rho(0)=1/2$ is the well-known intercept of the $\rho$-trajectory; $\alpha_B(0)\simeq -0.5$
is the intercept of the baryon trajectory, $\alpha_\Upsilon(0))$ is the intercept  of the 
$\Upsilon$- Regge trajectory, its value also has an uncertainty. Assuming its linearity, 
one can get $\alpha_\Upsilon(0))=-8, -16$ \cite{Piskunova}, 
while for the nonlinear ($b{\bar b}$) Regge trajectory
$\alpha_\Upsilon(0)\simeq 0$ \cite{Sergeenko:94,Likhoded:06}.
Inserting these values in the form for $f_{c({\bar c})}^{(n)}(x)$ and $f_{b({\bar b})}^{(n)}(x)$ 
we get the large sensitivity
for the $c$ and $b$ sea quark distributions in the $n$th Pomeron chain.
Note that the FFs also depend on the parameters of these Regge trajectories. Therefore,
the knowledge of the intercepts and slopes of the heavy-meson Regge trajectories is
very important for the theoretical analysis of open charm and beauty production in
hadron processes.

Note that all the quark distributions obtained within the QGSM are different from the parton
distributions obtained within the perturbative QCD which are usually compared with the experimental
data on the deep inelastic lepton scattering (DIS) off protons. To match these two kinds of quark 
distributions one can apply the procedure suggested in \cite{Cap-Kaid}.
The quantities $g_{cn}$ or $g_{bn}$ entering into Eq.(\ref {def:fc}) and Eq.(\ref {def:fb}) are 
replaced by the following new quantities depending on $Q^2$
\begin{equation}
{\tilde g}_{cn}=g_{cn}(1+\frac{Q^2}{Q^2+c})~;~ {\tilde g}_{bn}=g_{bn}\left(1+\frac{Q^2}{Q^2+d}\right)
\end{equation}
The parameters $c$ and $d$ are chosen such that the structure function constructed from the
valence and sea quark (antiquark) distributions in the proton should be the same as 
the one at the initial conditions at $Q^2=Q_0^2$ for the perturbative QCD evolution. A similar
procedure can be used to get the $Q^2$ dependence for the powers $a_{cn}$ and $a_{bn}$ entering
into Eqs.(\ref{def:fc},\ref{def:fb}) \cite{Cap-Kaid}. Then, using the DGLAP evolution equation 
\cite{DGLAP}, we can obtain the structure functions at large $Q^2$. 

\subsection{Fragmentation functions of partons to heavy flavour baryons }

 According to \cite{kaid2}, the production of $B$- baryons, for example 
$\Lambda,\Lambda_c,\Lambda_b$, is illustrated in Fig.2.   
Fig.2a illustrates the $B$ production in the fragmentation region, when the diquark in the 
colliding proton picks up directly the $b$ quark from the $q{\bar q}$ chain created between 
the diquark in the first colliding proton and the quark in the second proton. 
Fig.2b also illustrates the production of $B$ in the fragmentation region, when one quark 
and the string junction in the first proton pick up a quark ($u$ or $d$) and the $b$ quark from 
the $q{\bar q}$ chain mentioned above.
Figure.2c corresponds to the baryon-antibaryon $(B{\bar B})$ pair production in the central region.

\begin{figure}[htb]
\vspace{9pt}
\begin{center}
\includegraphics[scale=1.2]{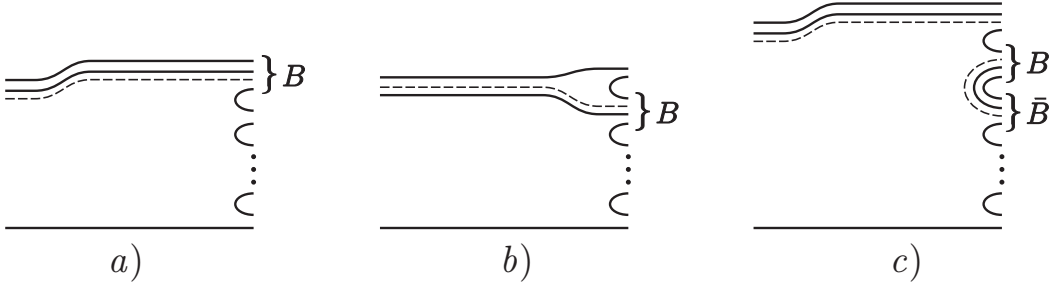}
\caption{Diagrams corresponding to the fragmentation of diquarks into baryons $B(\Lambda,\Lambda_c,\Lambda_b)$
in the fragmentation region (a) and in the central region (b), when the baryon-antibaryon pair ($B{\bar B}$)
is produced, the dashed line corresponds to the string junction between the diquark and the quark.}
\label{Fig.2}
\end{center}
\end{figure}

The FF of quark (diquark) chains to heavy flavor hadrons 
were constructed within the QGSM in \cite{Kaid-Pisk:1985,Piskunova}.
For example, the FF of quarks (antiquarks) to the charm and bottom baryons $\Lambda_c, \Lambda_b$
have the following forms \cite{Piskunova,Shabelski:95}:
\be
G_u^{\Lambda_c(\Lambda_b)}(z)=G_d^{\Lambda_c(\Lambda_b)}(z)=
a_0^{\Lambda_c(\Lambda_b)}(1-z)^{-\alpha_{\psi(\Upsilon)}(0)+\lambda+
2(\alpha_R(0)-\alpha_N(0))}
\label{def:Gu}
\ee
and
\be
G_{\bar u}^{\Lambda_c(\Lambda_b)}(z)=G_{\bar d}^{\Lambda_c(\Lambda_b)}(z)=
G_u^{{\bar\Lambda}_c({\bar\Lambda}_b)}=G_d^{{\bar\Lambda}_c({\bar\Lambda}_b)}=
G_u^{\Lambda_c(\Lambda_b)}(z)(1-z)^{2[1-\alpha_R(0)]}
\label{def:Gubar}
\ee
The FF of diquarks to $\Lambda_c.\Lambda_b$ are the sums of two parts, central 
$G_{0qq}^{\Lambda_c(\Lambda_)}$ and fragmentation $G_{fqq}^{\Lambda_c(\Lambda_)}$.
These parts have the following forms \cite{Shabelski:95}:
\be
G_{0uu}^{\Lambda_c(\Lambda_b)}=
a_0^{\Lambda_c(\Lambda_b)}
(1-z)^{-\alpha_{\psi(\Upsilon)}(0)+\lambda
-4[1-\alpha_N(0)]}
\label{def:Guuz}
\ee
and
\be
G_{fuu}^{\Lambda_c(\Lambda_b)}=
a_f^{\Lambda_c(\Lambda_b)}z^2
(1-z)^{-\alpha_{\psi(\Upsilon)}(0)+\lambda+1}~,
\label{def:Guuf}
\ee 
\be
G_{fud}^{\Lambda_c(\Lambda_b)}=G_{0ud}^{\Lambda_c(\Lambda_b)}=
a_f^{\Lambda_c(\Lambda_b)}z^{-2(\alpha_N(0)-\alpha_R(0)}
(1-z)^{-\alpha_{\psi(\Upsilon)}(0)+\lambda}~,
\label{def:Gudf}
\ee 
where $\lambda=2\alpha^\prime(0)_R<p_t^2>\simeq 0.5$.
To calculate inclusive spectra of $\Lambda_c$ or $\Lambda_b$ 
baryons we have to know the constants $a_0^{\Lambda_c(\Lambda_b)}$ and $a_f^{\Lambda_c(\Lambda_b)}$.
The constants $a_0^{\Lambda_c}$ and $a_f^{\Lambda_c}$ are presented in 
\cite{Piskunova}, their application allowed describing the experimental data on
the $\Lambda_c$ production. However, the similar constants for the $\Lambda_b$ production
are unknown. The question arises how to find these values.
The constant $a_f^{\Lambda_c}$ at $\alpha_\psi(0)=0$ was taken in \cite{Piskunova} as the same value 
as for the $\Lambda$ hyperon production $a_f^{\Lambda}$ \cite{kaid2}, e.g.,
$a_f^{\Lambda_c}=a_f^{\Lambda}=0.02$ assuming that in the fragmentation region (Fig.2(a,b)) the production 
mechanism for $\Lambda$ is the same as for the $\Lambda_c$ production. We can assume it for the 
$\Lambda_b$ production as well, e.g., $a_f^{\Lambda_b}=a_f^{\Lambda_c}=a_f^{\Lambda}=0.02$. 
As for the $\Lambda_b$ production in the central region, when
the $\Lambda_b$-${\bar\Lambda}_b$ pair is produced, see Fig.2c, one can estimate $a_0^{\Lambda_b}$
from the relation of the $\Lambda_b$ production in the central region to the $\Lambda_c$ production.
In the central region (Fig.2c), in addition to the production of two pairs of light quarks-antiquarks 
($u,d$), the pairs of heavy quarks-antiquarks ($(s{\bar s}), (c{\bar c}), (b{\bar b})$) are produced, 
when there are $(\Lambda{\bar \Lambda}),(\Lambda_c{\bar \Lambda}_c)$ and $(\Lambda_b{\bar \Lambda}_b)$ 
pairs in the final state.
Knowing the probability $w_{q{\bar q}}$ for the $(q{\bar q})$ pair production from the decay of the 
string between the diquark and the quark (Fig.2) one can relate $a_0^{\Lambda_c}$ to $a_0^{\Lambda}$
and $a_0^{\Lambda_b}$ to $a_0^{\Lambda_c}$, e.g.,
\be
a_0^{\Lambda_c}=\frac{w_{c{\bar c}}}{w_{s{\bar s}}}a_0^{\Lambda}     
\label{def:azeroc}
\ee
and 
\be
a_0^{\Lambda_b}=\frac{w_{b{\bar b}}}{w_{c{\bar c}}}a_0^{\Lambda_c}=
\frac{w_{b{\bar b}}}{w_{s{\bar s}}}a_0^{\Lambda}   
\label{def:azerob}
\ee
The probability of the $q{\bar q}$ pair production per unit of the four-dimension volume of the color tube 
$w_{q{\bar q}}$ was calculated in \cite{Gurvich} within the colour-tube fragmentation model 
\cite{Casher1,Gurvich} in the following form:
\be
w_{q{\bar q}}=\frac{\rho^2}{4\pi^3}exp(-\pi m_q^2/\rho)~,
\label{def:wqq}
\ee
where $\rho=1/(2\pi\alpha_R^\prime)$ is the so-called string tension, $\alpha_R^\prime$
is the slope of the Regge trajectory, $m_q$ is the quark mass. Note that Eq.(14) was obtained in
\cite{Gurvich} on the assumption of the linearity of the Regge trajectories.
Calculating $w_{c{\bar c}}$ and $w_{s{\bar s}}$ using Eq.(\ref{def:wqq}) and taking the value 
$a_0^{\Lambda}=0.3$ from \cite{kaid2} one can get $a_0^{\Lambda_c}=4\cdot 10^{-4}$ that coincides with
the value presented in \cite{Piskunova} which was used both at $\alpha_\Psi(0)=-2.18$ and $\alpha_\Psi(0)=0$
to describe rather successfully the experimental data on the $\Lambda_c$ production in $pp$ collisions. 
It makes us confident that one can use the same method to estimate $a_0^{\Lambda_b}$ as well.
Taking $\alpha_\Upsilon^\prime(0)\simeq 0.1$ (GeV$/c$)$^{-2}$ from \cite{Piskunova} and calculating
Eq.(\ref{def:azerob},\ref{def:wqq}) one can get  $a_0^{\Lambda_b}\simeq 1.7\cdot 10^{-6}$, which we  also
use both at $\alpha_\Upsilon(0)=-8$ and $\alpha_\Upsilon(0)=0$.

 \subsection{Charmed and beauty baryon production in $pp$ collision} 

The information on the charmonium ($c{\bar c}$) and bottomonium ($b{\bar b}$) Regge trajectories
can be found from the experimental data on the charmed and beauty baryon production in $pp$ 
collisions at high energies. For example, Fig.3 illustrates the sensitivity of the inclusive
spectrum $d\sigma/dx$ of the produced charmed  baryons $\Lambda_c$ to different values for 
$\alpha_\psi(0)$. 
The solid line corresponds to $\alpha_\psi(0)=0$, and the dashed curve corresponds to 
$\alpha_\psi(0)=-2.18$.
 Measuring the decay $\Lambda_c\rightarrow\Lambda^0 3\pi$,
the $R608$ experiment \cite{R608} at $\mid x\mid > 0.5$ and $\sqrt{s}=62$ GeV obtained
the $pp\rightarrow\Lambda_c X$ cross section $2.84\pm0.50\pm0.72 \mu$b.
The branching ratio of the decay 
$\Lambda_c\rightarrow\Lambda^0 3\pi$ is $2.8\pm 0.7\pm 1.1\%$ therefore, the cross section 
of the $\Lambda_c$ production is 
$\sigma(\mid x\mid > 0.5)=101\pm 18\pm 26 \mu$b. Theoretical expectation for this cross section
is  $\sigma(\mid x\mid > 0.5)=87.3\mu$b at $\alpha_\Psi(0)=0$ and $\sigma(\mid x\mid > 0.5)=30.5\mu$b
at $\alpha_\Psi(0)=-2.18$. On the other hand, the R422 experiment \cite{R422} was measuring the cross section of the
process $p\rightarrow e^- \Lambda_c X$ at $\mid x\mid>0.35$ with a large uncertainty from ($26\pm 12$) $\mu$b
to ($225\pm 9$) $\mu$b depending on the assumption made, see Table 6 in \cite{R422}.
So, the open circles in Fig.3 corresponding to the R608 experiment are more adequate to our calculations 
at $\alpha_\Psi(0)=0$, see the solid line in Fig.3.

\begin{figure}[ht]
   \begin{center}
 {\epsfig{file=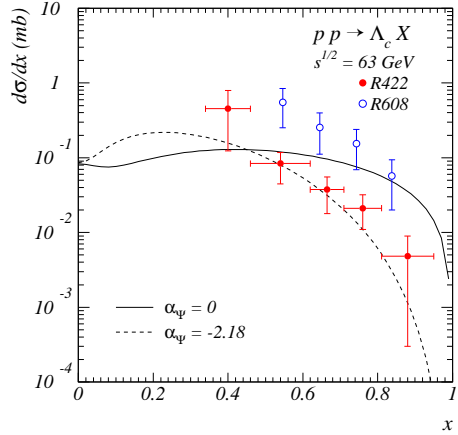,width=0.45\linewidth  }}
\caption[Fig.3]{The differential cross section $d\sigma/dx$ for the
 inclusive process $pp\rightarrow\Lambda_c X$ at $\sqrt{s}=62~\mathrm{GeV}$.
 The solid line corresponds to $\alpha_\psi(0)=0$, and the dashed curve corresponds to 
$\alpha_\psi(0)=-2.18$. The open circles correspond to the R608 experiment \cite{R608},
 and the dark circles correspond to the R422 experiment \cite{R422}.} 
\end{center}
 \end{figure}

High sensitivity of the inclusive spectrum $d\sigma/dx$ of the produced beauty baryons 
$\Lambda_b$ to different values for $\alpha_\Upsilon(0)$ is presented in Fig.4 (left).
\begin{figure}[htb]
\begin{center}
\begin{tabular}{cc}
\mbox{\epsfig{file=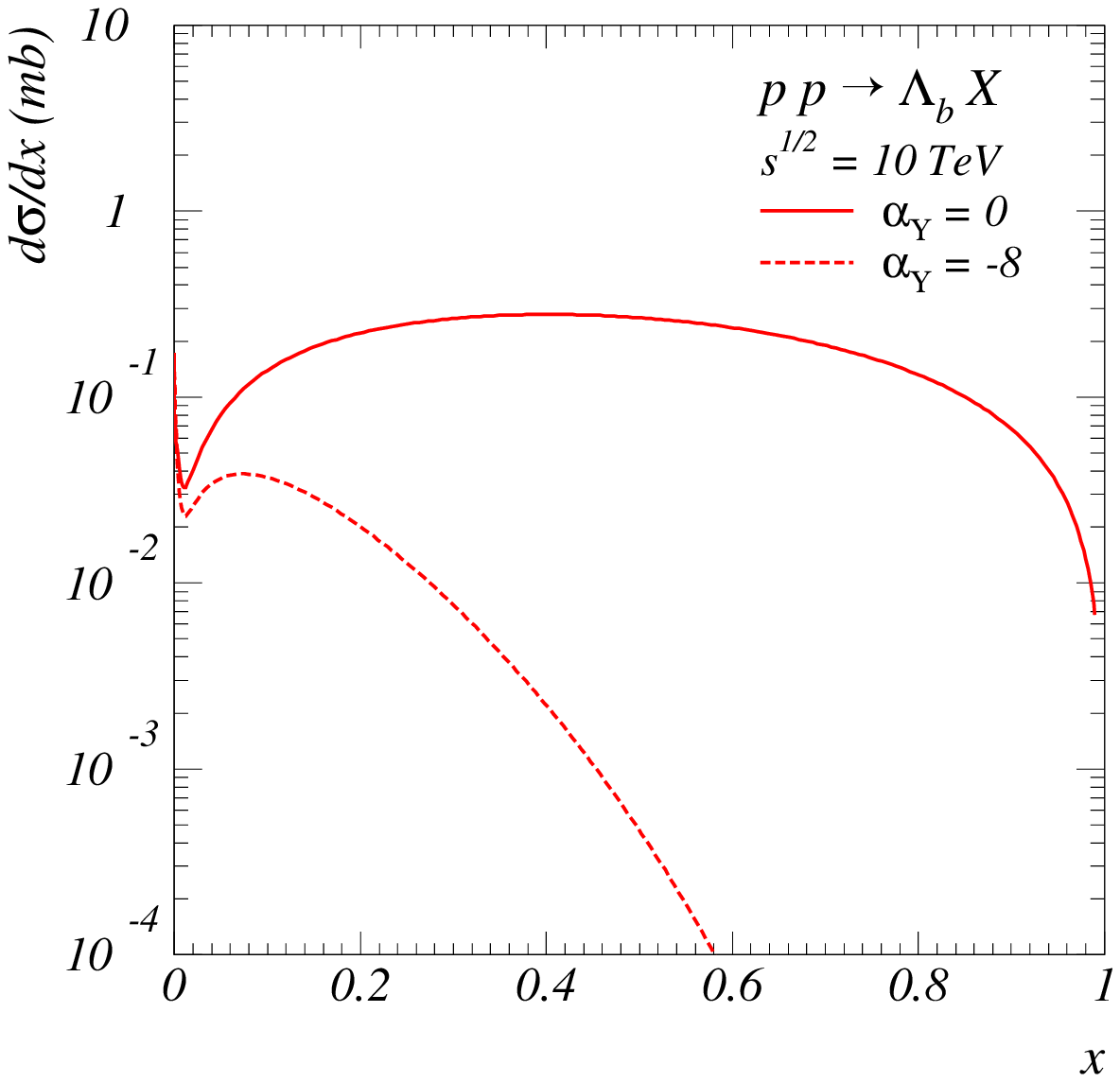,width=0.45\linewidth}} &
\mbox{\epsfig{file=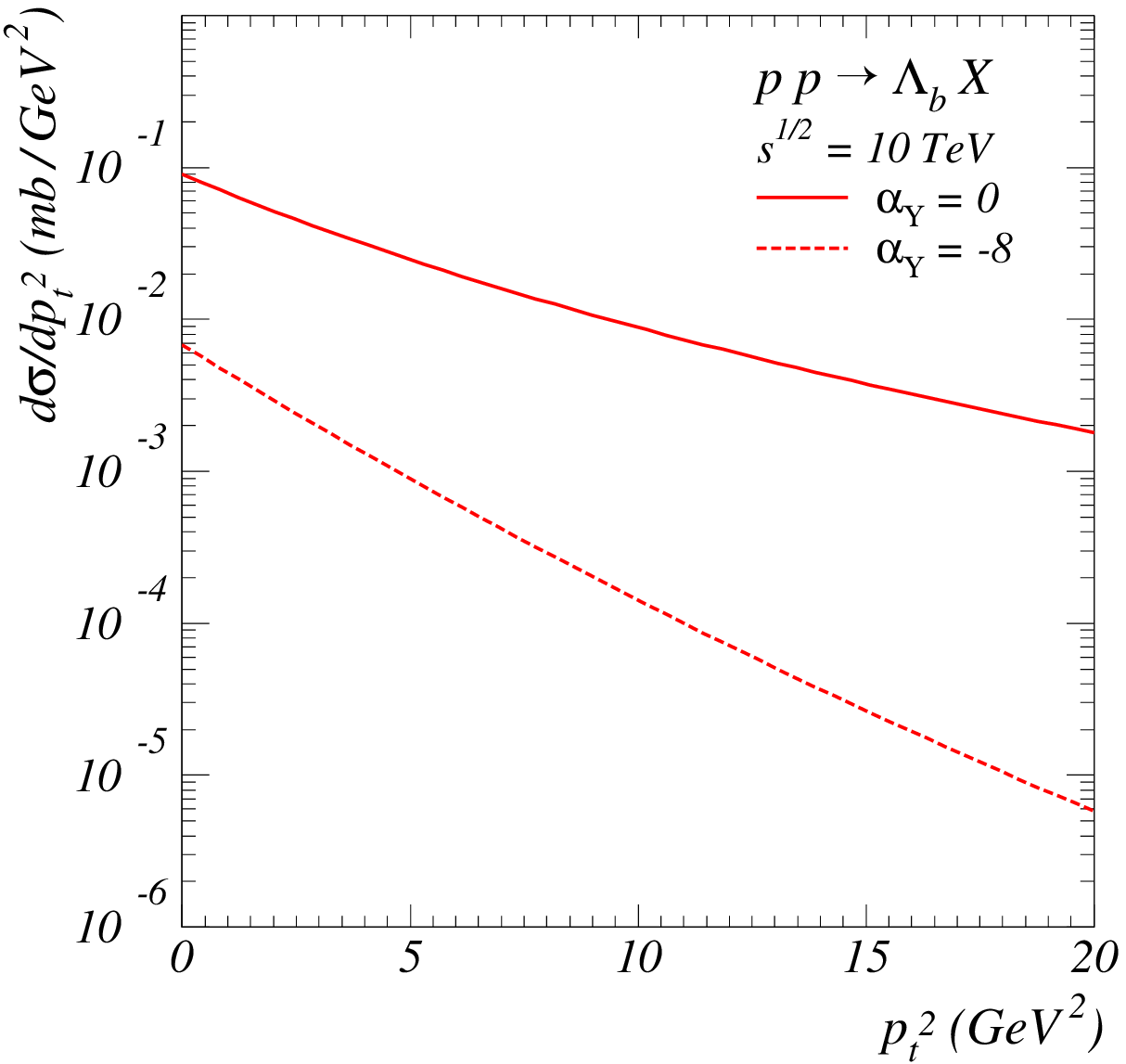,width=0.45\linewidth}}
\end{tabular}
\end{center}
 \caption[Fig.4\cite{R422}]{The differential cross section $d\sigma/dx$ (left) and
$d\sigma/dP_t^2$ (right) for the
 inclusive process $pp\rightarrow\Lambda_b X$ at $\sqrt{s}=10~\mathrm{TeV}$.
}
\end{figure}
The $p_t$-inclusive spectrum of $\Lambda_b$ has much lower sensitivity to this quantity,
according to the results presented in Fig.4 (right).
Actually, our results presented in Fig.4 could be considered as some predictions for 
future experiments at LHC.

The produced $\Lambda_b$ baryon can decay $\Lambda_b\rightarrow J/\Psi \Lambda^0$
 with the branching ratio $Br_{\Lambda_b\rightarrow J/\Psi\Lambda^0}=
\Gamma_{\Lambda_b\rightarrow J/\Psi\Lambda^0}/\Gamma_{tot}=(4.7\pm 2.8)\cdot 10^{-4}$ 
and $J/\Psi$ decays into $\mu^+\mu^-$ ($Br_{J/\Psi\rightarrow\mu^+\mu_-}=(5.93\pm 0.06)\%$)
 or into $e^+e^-$ ($Br_{J/\Psi\rightarrow e^+e^-}=5.93\pm 0.06\%$), whereas $\Lambda^0$ can decay into 
$p\pi^-$ ($Br_{\Lambda^0\rightarrow p\pi^-}=63.9\pm 05\%$), or into $n\pi^0$ 
($Br_{\Lambda^0\rightarrow n\pi^0}=35.8\pm 0.5\%$).
Experimentally one can measure the differential cross section
$d\sigma/d\xi_p dt_p dM_{J/\Psi}$, where $\xi_p=\Delta p/p$ is the energy loss, $t_p=(p_{in}-p_1)^2$ is 
the four-momentum transfer, $M_{J/\Psi}$ is the effective mass of the $J/\Psi$-meson. 
\begin{figure}[htb]
\begin{center}
\begin{tabular}{cc}
\mbox{\epsfig{file=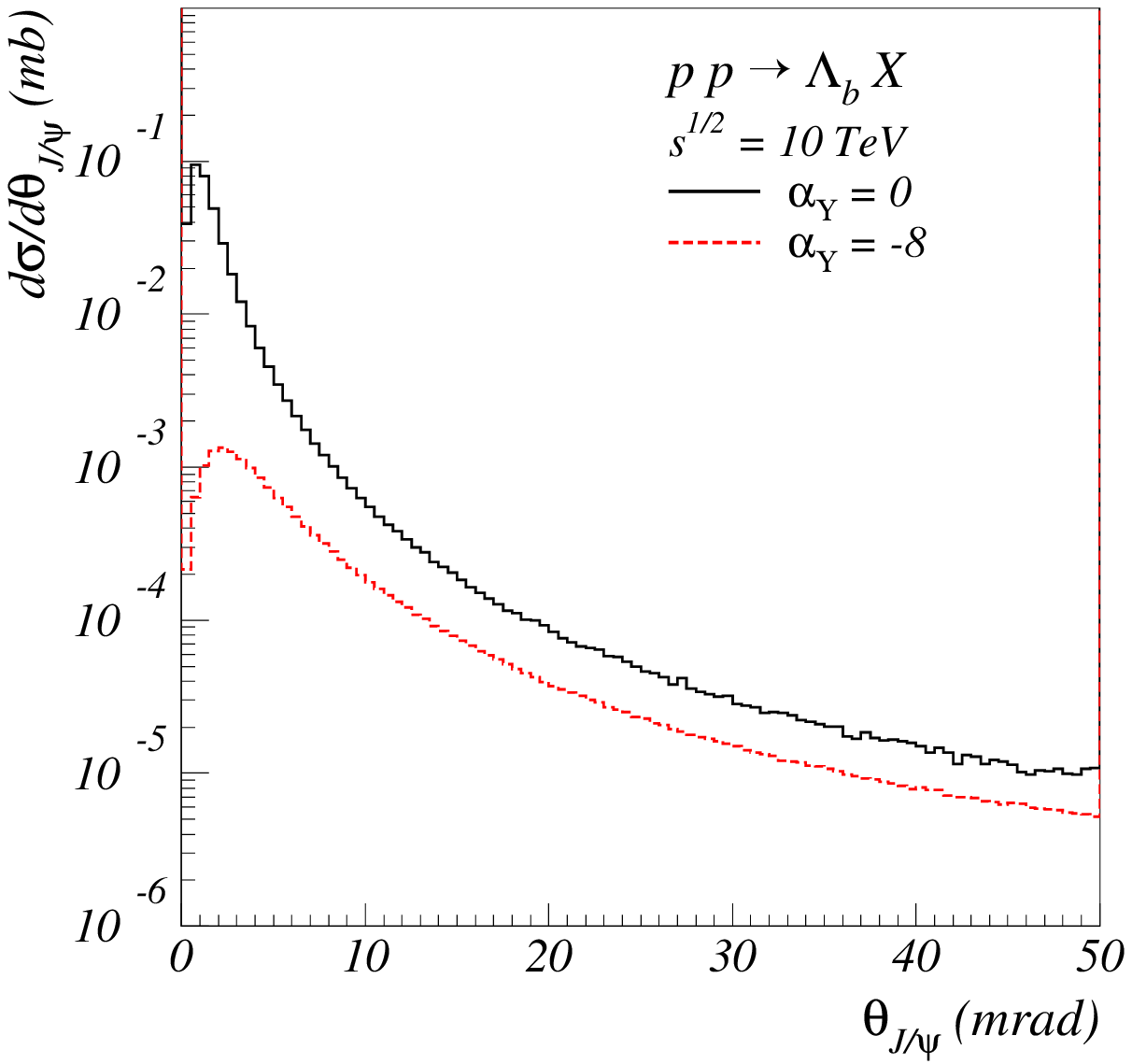,width=0.45\linewidth}} &
\mbox{\epsfig{file=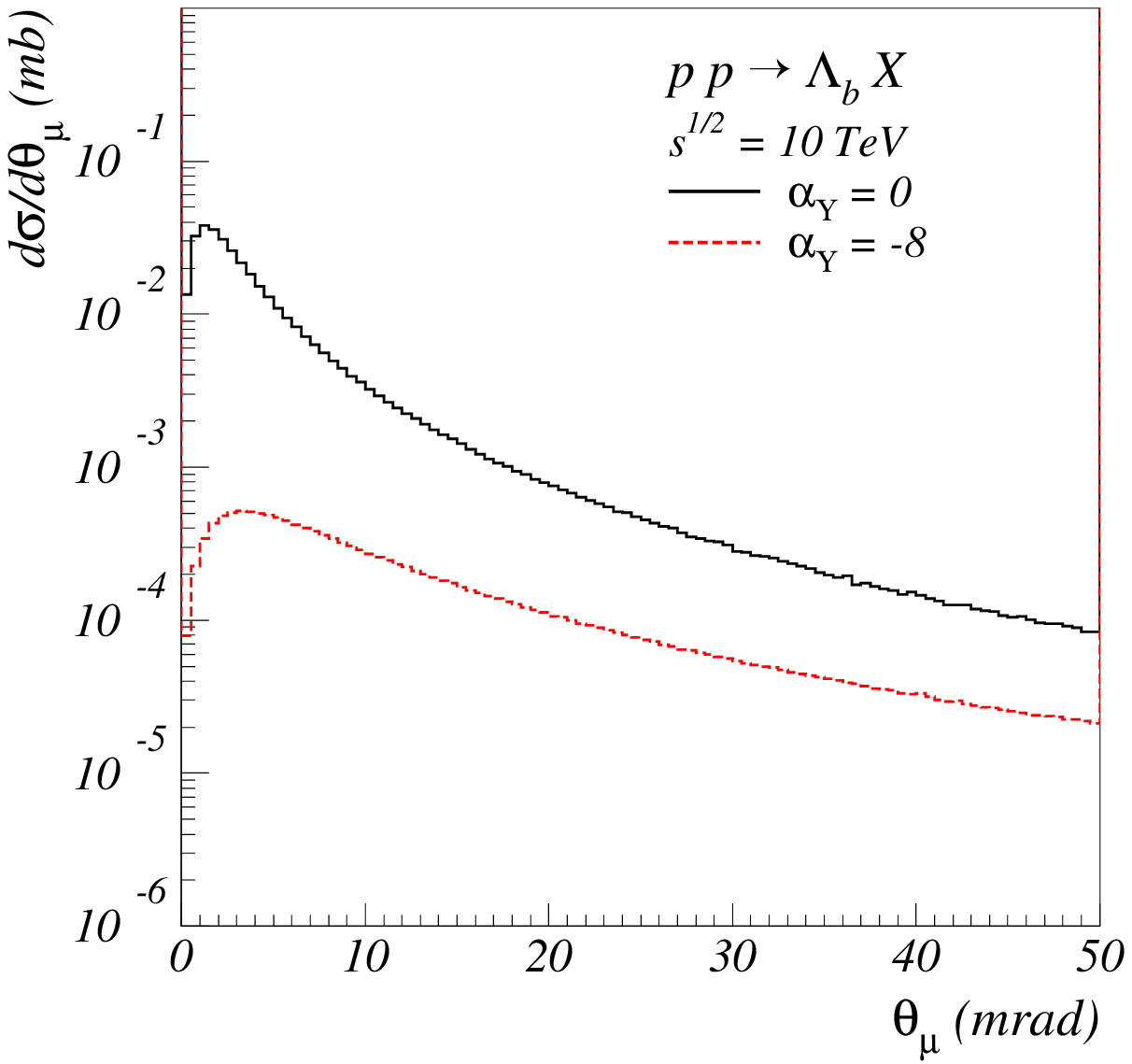,width=0.45\linewidth}}
\end{tabular}
\end{center}
 \caption[Fig.5]{The distributions over $\theta_{J/\Psi}$ (left) and $\theta_{\mu^+}$
(right) 
for the inclusive process 
$pp\rightarrow\Lambda_b X\rightarrow\mu^+\mu^- p\pi^- X$ at $\sqrt{s}=10$ TeV,
where $\theta_{J/\Psi}$ and $\theta_{\mu^+}$ are the scattering angles for $J/\Psi$
and $\mu^+$.
}
\end{figure}

\begin{figure}[htb]
\begin{center}
\begin{tabular}{cc}
\mbox{\epsfig{file=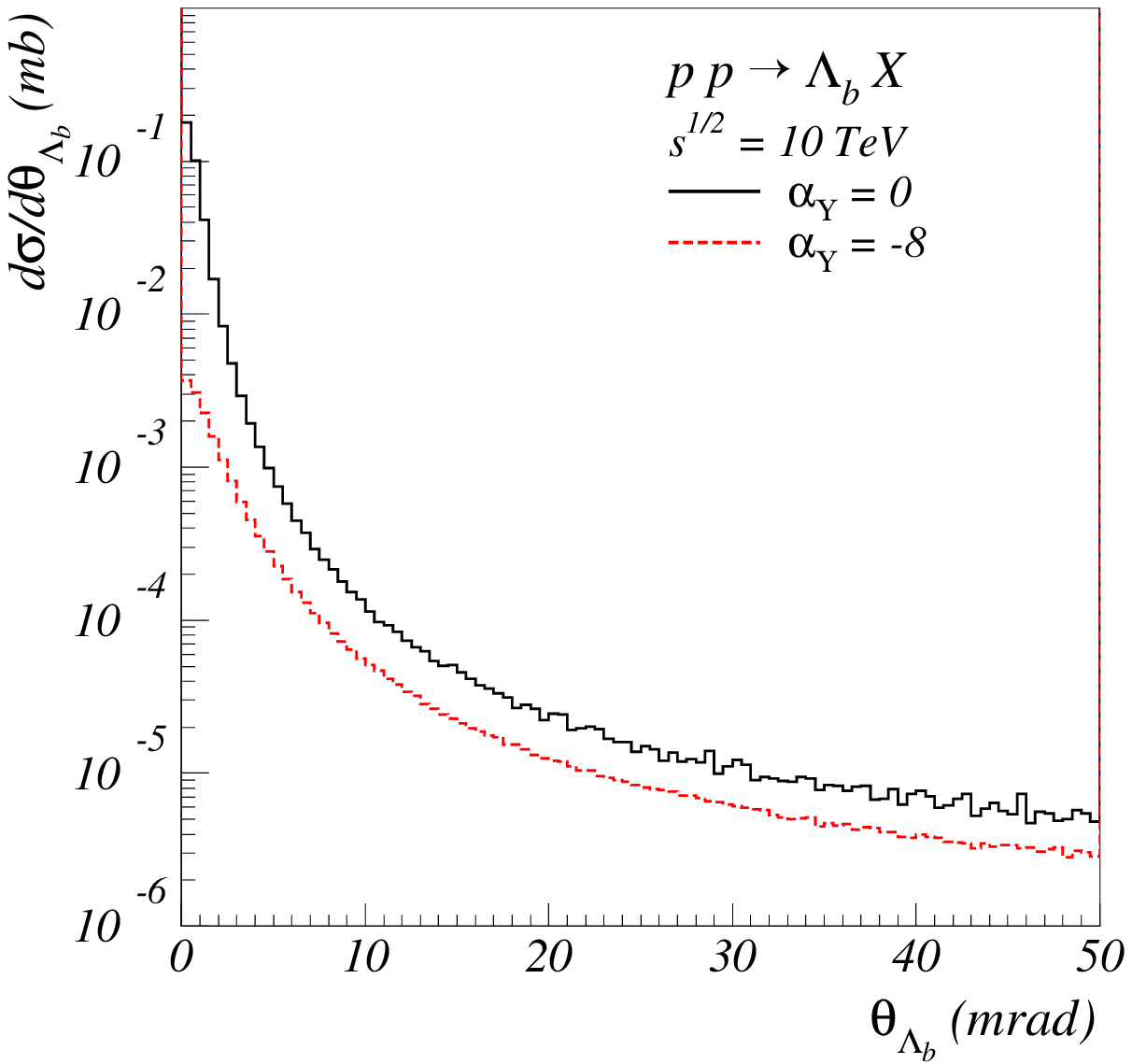,width=0.45\linewidth}} &
\mbox{\epsfig{file=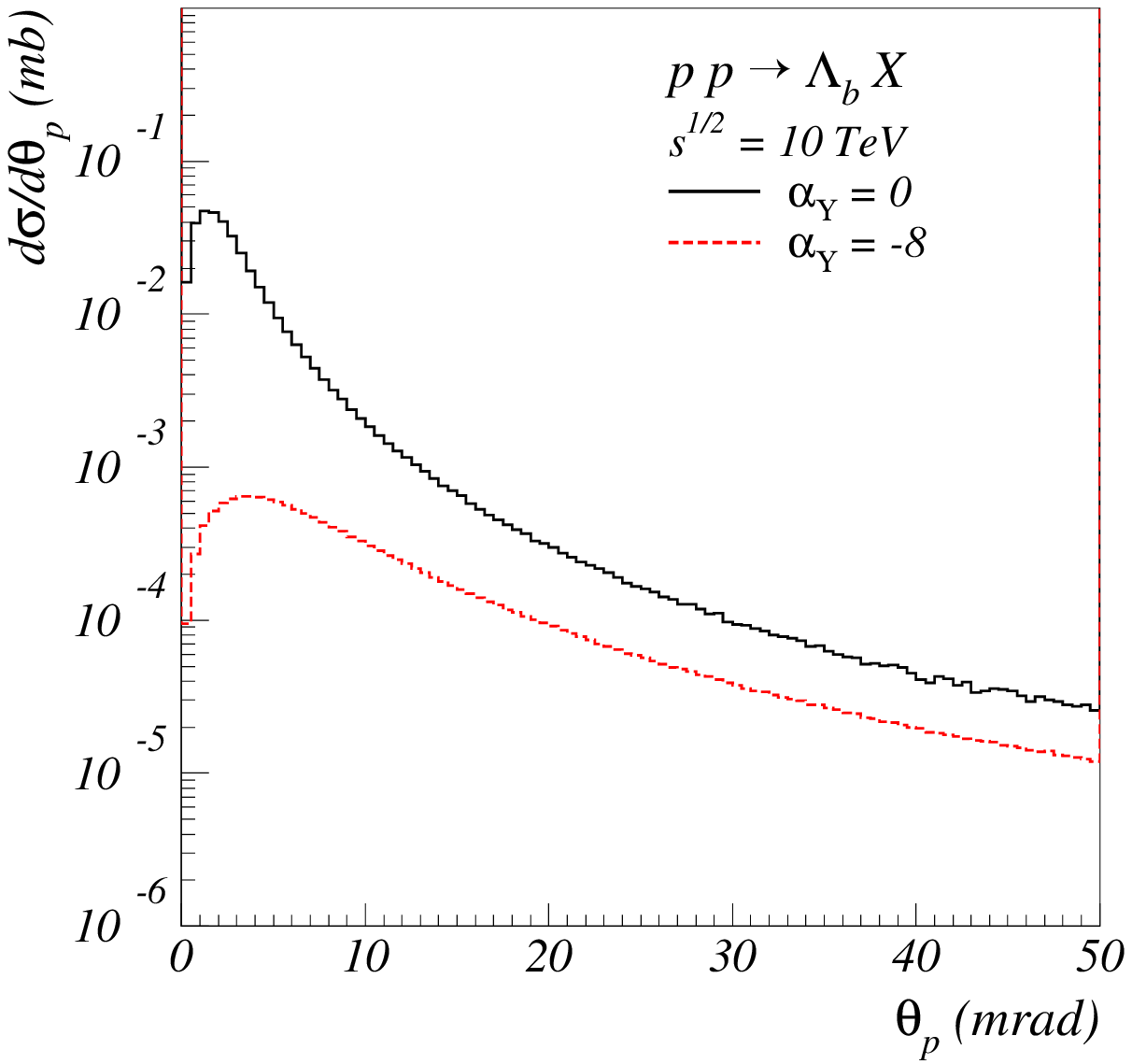,width=0.45\linewidth}}
\end{tabular}
\end{center}
 \caption[Fig.6]{The distributions over $\theta_{\Lambda_b}$ (left) and $\theta_p$ (right) 
for the inclusive process 
$pp\rightarrow\Lambda_b X\rightarrow\mu^+\mu^- p\pi^- X$ at $\sqrt{s}=10$ TeV.}
\end{figure}
\begin{figure}[htb]
\begin{center}
\begin{tabular}{cc}
\mbox{\epsfig{file=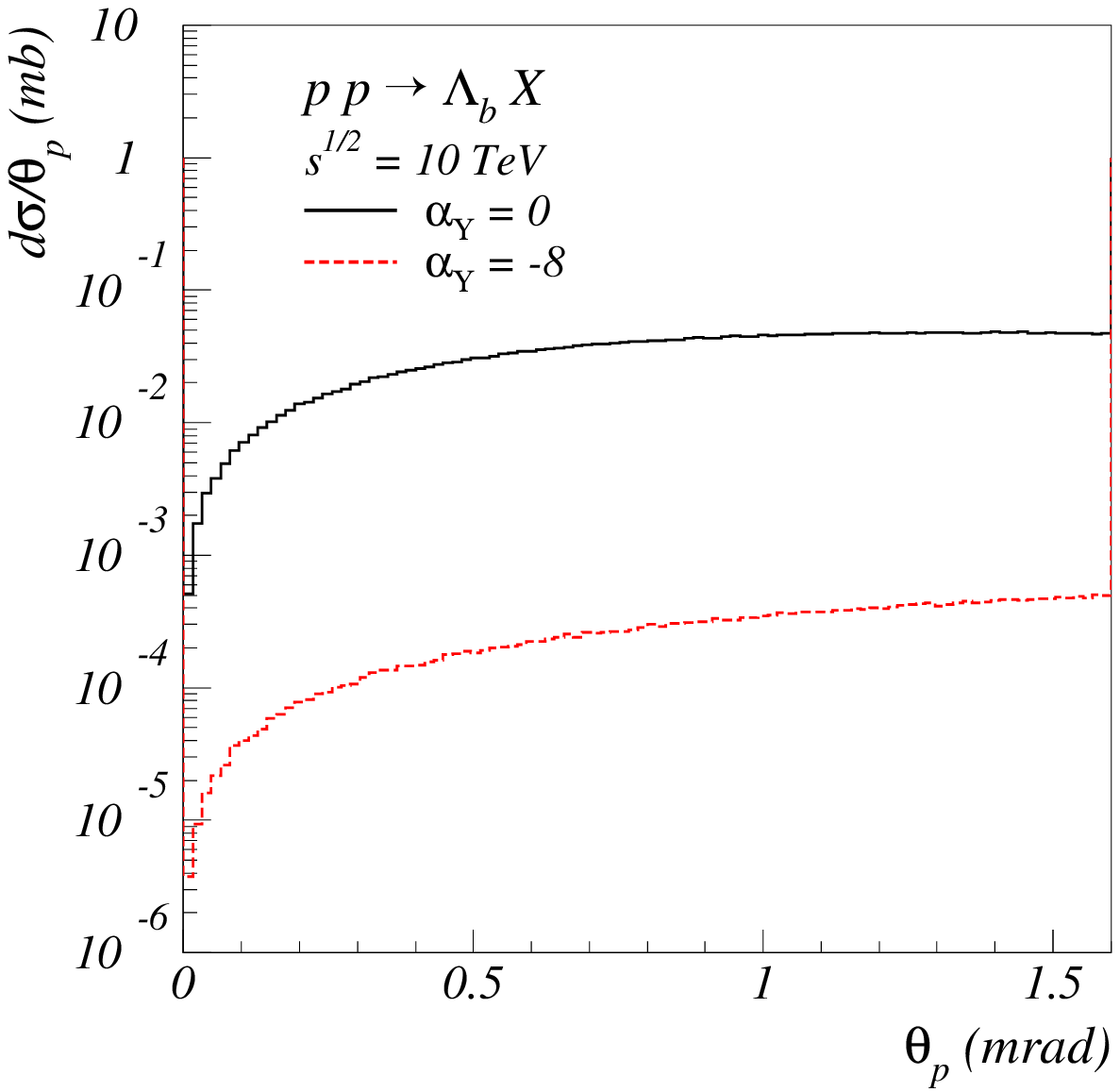,width=0.45\linewidth}} &
\mbox{\epsfig{file=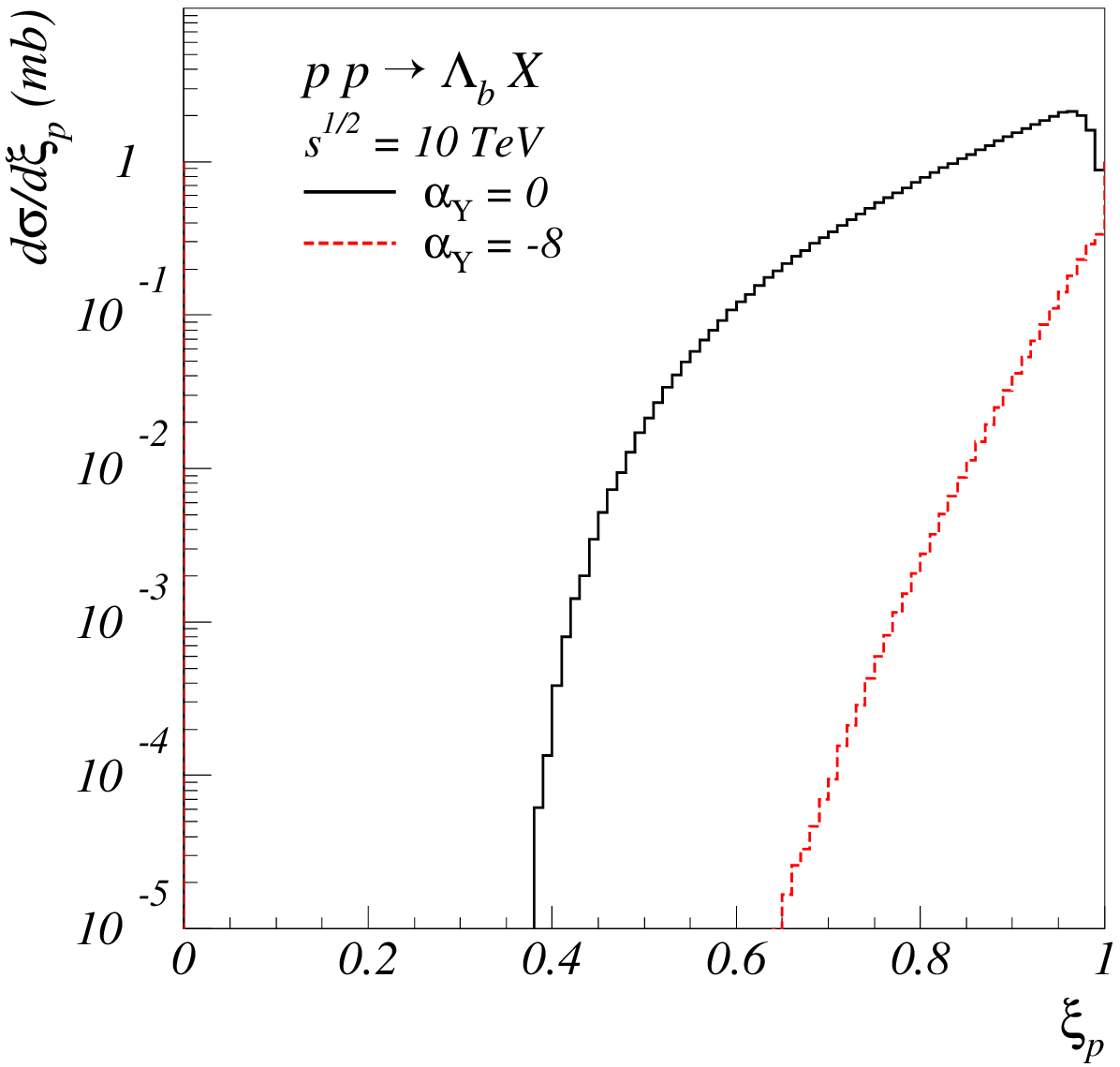,width=0.45\linewidth}}
\end{tabular}
\end{center}
 \caption[Fig.7]{The distributions over $\theta_p$ for the inclusive process 
$pp\rightarrow\Lambda_b X\rightarrow\mu^+\mu^- p\pi^- X$ at $\sqrt{s}=10$ TeV. 
The solid curve corresponds to $\alpha_\Upsilon(0)=0$; the long dashed line corresponds
to $\alpha_\Upsilon(0)=-8$ (left).
Right distributions over $\xi_p=(\sqrt{s}-2E_p)/\sqrt{s}$ for the inclusive process 
$pp\rightarrow\Lambda_b X\rightarrow\mu^+\mu^- p\pi^- X$ at $\sqrt{s}=10.$ TeV.}
\end{figure}

\section{Results and discussions}
Now let us analyze the production of the beauty hyperon, namely $\Lambda^0_b$,
at small scattering angles $\theta_{\Lambda_b}$ in the $pp$ collision at LHC energies.
This study would be reliable for the forward experiments at the LHC.
In Fig.5 the  distributions over $\theta_{J/\Psi}$ (left) and $\theta_{\mu^+}$ (right) are presented 
at different values of the intercept $\alpha_\Upsilon(0)=0$ (solid line) and 
$\alpha_\Upsilon(0)=-8$ (dashed line), 
where $\theta_{J/\Psi}$ is the scattering angle for the final $J/\Psi$. 
Figure 5 shows a large sensitivity of these distributions to the intercept $\alpha_\Upsilon=0$ 
and $\alpha_\Upsilon=-8$ of the $\Upsilon(b{\bar b})$ Regge trajectory at small values of the scattering angle.
It also shows that the angle distribution of the $\mu$ meson produced from the $J/\Psi$ decay is broader than
the one for the $J/\Psi$ meson.  

In Fig.6 the  distributions over $\theta_{\Lambda_b}$ (left) and $\theta_p$ (right) are presented 
at different values of the intercept $\alpha_\Upsilon(0)=0$ (solid line) and
$\alpha_\Upsilon(0)=-8$ (dashed line),
where $\theta_{\Lambda_b}$ is the scattering angle for the $\Lambda_b$ baryon and
$\theta_p$ is the scattering angle of the final proton.
One can see from Fig.6 that the angle distribution of protons produced from the $\Lambda^0$ decay is broader than
the one for the $\Lambda^0$ hyperon. 
In Fig.7 the same $\theta_p$ distribution as in Fig.6 is presented but at very low values, $\theta_p<1.6$ mrad, 
which can be observed in the TOTEM experiment at CERN \cite{TOTEM}.

All the angle distributions presented in Figs.5-7 show a huge sensitivity to the values of the intercepts of the
$\Upsilon(b{\bar b})$ Regge trajectory $\alpha_\Upsilon(0)=0$ and $\alpha_\Upsilon(0)=-8$ at the 
angles about $4-8$ mrad for muons, pions and protons and at the angles about $2-4$ mrad. for $\Lambda_b$ baryons 
and $J/\Psi$ vector mesons with the hidden charm. 
The distribution over the variable $\xi_p$ related to the proton loss energy presented in Fig.7 (right) shows that
the cross section  for $\alpha_\Upsilon(0)=-8$ is too low at $\xi_p<0.6$ and, in fact, it cannot be measured
experimentally.    

From the distributions in Figs. 5-7, the following experimental signatures
can be deduced.

The ATLAS forward detectors could register the decay
$\Lambda^0_b\rightarrow J/\Psi~\Lambda^0\rightarrow \mu^+\mu^-~\pi^0 n$
by detecting the neutron  in the ZDC~\cite{ZDC} and  
two muons in the muon detector. This possibility is under study.

The TOTEM~\cite{TOTEM} together with the CMS might be able to measure the channel
$\Lambda_b\rightarrow J/\Psi~\Lambda^0\rightarrow \mu^+\mu^-~\pi^- p$
(the integrated cross-section is about 0.2-0.3 $\mu$b at $\alpha_\Upsilon(0)=0$ and
smaller at $\alpha_\Upsilon(0)=-8$).
The T2 and T1 tracking stations of the TOTEM apparatus have their angular acceptance 
in the intervals $\rm 3\,mrad < \theta < 10\,mrad$ (corresponding to $6.5 > \eta > 5.3$) 
and $\rm 18\,mrad < \theta < 90\,mrad$ (corresponding to $4.7 > \eta > 3.1$) 
respectively, 
and could thus detect 42\% of the muons from the $J/\Psi$ decay (Fig. 5).
In the same angular intervals, 36\% of the $\pi^{-}$ and 35\% of 
the protons from the $\Lambda^0$ decay are expected. 
According to a very preliminary estimate~\cite{Deile}, protons with energies 
above 3.4\,TeV
emitted at angles smaller than 0.6\,mrad 
could be detected 
in the Roman Pot station at 147\,m from IP5 \cite{TOTEM,Deile}. In the latter case, 
the reconstruction of the proton kinematics may be possible, whereas 
the trackers T1 and T2 do not provide any momentum or energy information.
Future detailed studies are to establish the full event topologies with
all correlations between the observables in order to assess whether the 
signal events can be identified and separated from backgrounds. These 
investigations should also include the CMS calorimeters HF and CASTOR which 
cover the same angular ranges as T1 and T2 respectively \cite{Deile}.

\section{Conclusion}
 We analyzed the production of charmed and beauty baryons in proton-proton collisions at high energies 
within soft QCD, namely the quark-gluon string model (QGSM). This approach can describe 
rather satisfactorily the charmed baryon production in $pp$ collisions \cite{LAS}-\cite{LLB:09}. 
It allows us to apply the QGSM to studying the beauty baryon production in $pp$ collisions. 
We focus mainly on the analysis of the forward $\Lambda_b$ 
production in $pp$ collisions at LHC energies and got some predictions which could be reliable at the
TOTEM and ATLAS experiments at CERN. It is shown that information on the sea 
$b$-quark distributions in the proton and the fragmentation functions of quarks (diquarks) into beauty baryon
$\Lambda_b$ can be extracted from this analysis. 
We also show that the angle distributions of the hadrons produced in the reactions
$pp\rightarrow\Lambda_b X\rightarrow\mu^+\mu^- p\pi^- X$ or $pp\rightarrow\Lambda_b X\rightarrow\mu^+\mu^- n\pi^0 X$
calculated within the QGSM are very sensitive to the intercept values $\alpha_\Upsilon(0)=0, -8$ of the
$\Upsilon(b{\bar b})$- Regge trajectory. It means that the experiments on the forward $\Lambda_b$
production in $pp$ collisions at the LHC could give us the answer whether the $\Upsilon(b{\bar b})$- 
Regge trajectory is linear ($\alpha_\Upsilon(0)=-8$) or nonlinear ($\alpha_\Upsilon(0)=0$); the latter is predicted
by the perturbative QCD. We did not include the diffractive and double diffractive $\Lambda_b$ production
in $pp$ collisions because these processes can be experimentally
separated from the forward $\Lambda_b$ production
\cite{TOTEM,Deile}.      

Note  that in this paper we neglect the contribution of the intrinsic charm in the proton calculating the 
charmed baryon production in $pp$ collisions  
and possible intrinsic beauty in the proton. However, as shown 
recently \cite{Ullrich:2010}, the intrinsic charm in the proton can result in a sizable contribution to the forward 
charmed meson production. Therefore, we intend to include this effect in the next more detailed study of this problem.

\section{Acknowledgments}

 We are very grateful to M. Deile, P. Garfstr{\"o}m, and  N.I. Zimin    
for extremely useful help related to the possible experimental check of the
suggested predictions at the LHC and the preparation of this paper.
We also thank A.V. Efremov, K. Eggert, A. B. Kaidalov,  
A. D. Martin and M. Poghosyan for very useful discussions. 
This work was supported in part by the Russian Foundation for Basic Research 
grant N: 08-02-01003.


\begin{thebibliography}{99}\itemsep -1mm
%
\bibitem{Nasson}
P.~Nasson, S.~Dawson and R.`K.~Ellis, 
Nucl. Phys. B 303 607 (1988); {\it ibid.} 
 B 327 (1989) 49; {\it ibid.} 
 B 335 (1989) 260E.
\bibitem{Kniehl3}
B.~A.~Kniehl, G.~Kramer, I.~Schienbein and H.~Spiesberger,
Phys.Rev D 77 (2008) 014011; arXiv:075054392v.2 [hep-ph].
\bibitem{tHooft:1974}
G.~t'Hooft, Nucl.Phys. B72 (1974) 461.
\bibitem{Veneziano:1975}
M.~Ciafaloni, G.~Marchesini, G.~Veneziano,
Nucl.Phys. B 98 (1975) 472.
\bibitem{Casher1}
A.~Casher, J.~Kogut, L.`Susskind,
Phys.Rev., D 10 (9174) 732. 
\bibitem{Gurvich}
E.~G.~Gurvich,
Phys.Lett. B 87 (1979) 386.
\bibitem{kaid1}
A.~B.~Kaidalov, Phys. Lett. B 116 (1982) 459;
A.~B.~Kaidalov and K.~A.~Ter-Martirosyan,
Phys. Lett. B 117 (1982) 247.
\bibitem{Werner:1993}
K.~Werner,
Phys.Rep. 232 (1993) 87.
\bibitem{Capella:1994}
A.~Capella, U.~Sukhatme, C.~I.~Tan, J.~Tran Than Van,
Phys. Rep. 236 (1994) 225.
\bibitem{kaid2}
A.~B.~Kaidalov and O.~I.~Piskunova,
Z. Phys. C30 (1986) 145.
\bibitem{Ter-Mart}
K.~A.~Ter-Martirosyan, Phys. Lett. B44 (1973) 377. 
\bibitem{LAS}
G.~I.~Lykasov, G.~H.~Arakelian and M.~N.~Sergeenko,
Phys. Part. Nucl. 30 (1999) 343; G.I.~Lykasov,
M.~N.~Sergeenko, Z.Phys., C 52 (1991) 635.
\bibitem{LKSB:09}
G.~I.~Lykasov, Z.~M.~Karpova, M.~N.~Sergeenko and V.~A.~Bednyakov,
Europhys.Lett. 86 (2009) 61001; arXiv: 0812.3220 [hep-ph]. 
\bibitem{LLB:09}
G.~I.~Lykasov, V.~V.~Lyubushkin and V.~A.~Bednyakov, 
Nucl.Phys.(Proc.Suppl.) B 198(2010) 165; arXiv:0909.5061 [hep-ph]. 
\bibitem{Boresk-Kaid:1983}
K.~G.~Boreskov, A.~B.~Kaidalov,
Sov.J.Nucl.Phys. 37 (1983) 100.
\bibitem{Sergeenko:94}
M.~N.~Sergeenko,
Z.Phys. C 64 (1994) 315.
\bibitem{Likhoded:06}
S.~S.~Gershtein, A.~K.~Likhoded and A.~V.~Luchinsky,
Phys.Atom.Nucl. 70 (2006) 759.
\bibitem{Cap-Kaid}
A.~Capella, A.~B.~Kaidalov, C.~Merino and J.~Tran Than Van,
Phys.Lett. B 337 (1994) 358; {\it ibid} ~ B 343  (1995) 403.
\bibitem{DGLAP}
V.~N.~Gribov and L.~N.~Lipatov, Sov.J.Nucl.Phys. 15 (1972) 438;
G.~Altarelli and G.~Parisi, Nucl.Phys. B 126 (1997) 298;
Yu.L.~Dokshitzer, Sov.Phys. JETP 46 (1977) 641.
\bibitem{Kaid-Pisk:1985}
A.~B.~Kaidalov, O.~I.~Piskunova,
Sov.J.Nucl.Phys. 41 (1986) 1278;
{\it ibid}
Sov.J.Nucl.Phys. 43 (1986) 994.
\bibitem{Piskunova}
O.~I~ Piskunova,  
Yad. Fiz. 56 176 (1993) (Phys.Atom.Nucl. 56 (1993) 1094;
{\it ibid} 64  (2001) 392.
\bibitem{Shabelski:95}
N.~Armesto, C.~Pajares and Y.~M.~Shabelski,
arxiV:9506212 [hep-ph].
\bibitem{R608}
P.~Chauvat, et al., Phys.Lett., B 199 (1987)304.
\bibitem{R422}
G.~Bari, et al., Nuovo Cim., A 104 (1991) 571.
\bibitem{ZDC}
ATLAS Collaboration, Letter of Intent ``Zero Degree Calorimeters
for ATLAS'', CERN/LHCC/2007-001, LHCC I-013, 12 January, 2007.
\bibitem{TOTEM} TOTEM: Technical Design Report, CERN-LHCC-2004-002 (2004); 
addendum CERN-LHCC-2004-020. G.~Anelli {\it et al.}  [TOTEM Collaboration],
``The Totem Experiment At The CERN Large Hadron Collider'', JINST {\bf 3}, 
S08007 (2008).
\bibitem{Deile} 
M.~Deile, private communication, and
H.~Niewiadomski, TOTEM-NOTE-2009-002.
\bibitem{Ullrich:2010}
V.~P.~Goncalves, F.~S.~Navarra, T.~Ullrich, 
arXiv:0805.0810 [hep-ph]. 
\end{thebibliography}
\end{document}